\documentstyle[twoside,fleqn,espcrc2,epsfig]{article}

\newcommand{\AmS}{{\protect\the\textfont2
  A\kern-.1667em\lower.5ex\hbox{M}\kern-.125emS}}

\newcommand{\be}{\begin{equation}}
\newcommand{\ee}{\end{equation}}
\newcommand{\ba}{\begin{eqnarray}}
\newcommand{\ea}{\end{eqnarray}}

\newcommand{\ga}{\gamma_5} 
\newcommand{\dg}{^{\dagger}}
\newcommand{\la}{\lambda}
\newcommand{\re}[1]{(\ref{#1})}
\newcommand{\F}{{\cal F}} 
\newcommand{\df}{[\mbox{d}\bar{\psi}\mbox{d}\psi]}
\newcommand{\f}{{\mbox{\tiny f}}} 
\newcommand{\e}{\mbox{e}}
\newcommand{\cao}{{\cal O}} 
\newcommand{\G}{\Gamma} 
\newcommand{\bG}{\bar{\Gamma}} 
\newcommand{\pt}{\partial} 
\newcommand{\Id}{\mbox{1\hspace{-0.98mm}l}}   

\title{\vspace*{-6mm}
\raisebox{0.8cm}[0pt][0pt]{\makebox[0pt][l]{\parbox{16cm}{\normalsize%
\mbox{}\hfill HUB-EP-99/38}}}\\
Dirac operator normality and chiral properties\thanks{Talk
presented at Lattice '99, Pisa, Italy.}}

\author{ W.~Kerler \address{Institut f\"ur Physik, 
Humboldt-Universit\"at, D-10115 Berlin, Germany}}

\begin{document}

\begin{abstract}
Normality and $\ga$-hermiticity are what gives rise to chiral properties and 
rules. The Ginsparg-Wilson (GW) relation is only one of the possible spectral
constraints. The sum rule for chiral differences of real modes has important
consequences. The alternative transformation of L\"uscher gives the same 
Ward identity as the usual chiral one (if zero modes are properly treated). 
Imposing normality on a general function of the hermitean Wilson-Dirac 
operator $H$ leads at the same time to the GW relation and to the Neuberger 
operator. 
\end{abstract}

\maketitle
\thispagestyle{empty}

\section{BASIC RELATIONS}

Normality of the Dirac operator $D$ 
\be
[D,D\dg]=0
\label{nor}
\ee         
implies that with $D f_k = \la_k f_k$ one also has $D\dg f_k = \la_k^* f_k$. 
This together with $\ga$-hermiticity
\be
D\dg = \ga D \ga  \, \, ,
\label{Ddg}
\ee     
by which one then has $D\ga f_k = \la_k^*\ga f_k$, leads to
\be
[\ga,D] f_k = 0 \quad \mbox{ if } \quad \la_k \mbox{ real } ,
\label{egc}
\ee
i.e. to simultaneous eigenvectors of $\ga$ and $D$ in the subspace of real 
eigenvalues of $D$. This exactly is the basis of chiral properties.

In addition to eigenvectors with $\ga f_k=\pm f_k$ for real $\la_k$, one obviously gets pairs of eigenvectors $f_k$ and $\ga f_k$ of $D$ with 
complex eigenvalues  $\la$ and $\la^*$ , respectively, so that one has the  
relation
\be
f_l\dg \ga f_k = 0 \quad \mbox{ for } \quad \la_l^* \ne \la_k \, \, .
\label{nela}
\ee

Normality of $D$ is necessary and sufficient in order that the eigenvectors 
form a complete set in unitary space (as one has on a finite lattice). 
With this, \re{nela}, and Tr$(\ga)=0$ we obtain
\be
\sum_{\la \mbox{ \scriptsize  real }} \Big(N_+(\la) - N_-(\la)\Big) = 0 
\label{res}
\ee
where $N_\pm(\la)$ is the number of modes with chirality $\pm 1$ for real 
eigenvalue $\la$ of $D$. The sum rule for the chiral differences of real modes
\re{res} has the remarkable consequence that $N_-(0) - N_+(0)$, 
the index of $D$, can only be nonvanishing if a corresponding contribution 
from nonzero $\la$ exists.  

Again using the completeness of the eigenvectors and \re{nela} we get the
relations 
\be
\varepsilon \mbox{Tr}((D+\varepsilon)^{-1}\ga) \rightarrow N_+(0) - N_-(0)
\label{re0}
\ee
\be
\mbox{Tr}((D+\varepsilon)^{-1}\ga D) \rightarrow 
\sum_{\la\ne 0 \atop \mbox{ \tiny  real }} \Big(N_+(\la) - N_-(\la)\Big)
\label{re1}
\ee
\vspace*{-3mm}
for $\varepsilon \rightarrow 0$ and also 
\be
\mbox{Tr}(\ga D)=
\sum_{\la\ne 0\mbox{ \scriptsize real}} \la\,\Big(N_+(\la) - N_-(\la)\Big)
\, \, .
\label{rel}
\ee

 From normality of $D$ it follows that in the decomposition 
$D = u + i v$ with
\be
u= \frac{1}{2} (D+D\dg) \, , \quad v= \frac{1}{2i} (D-D\dg)
\label{Duv}
\ee
the hermitean operators $u$ and $v$ commute. Therefore the eigenvalues of 
$u$ and $v$ are simply the real and imaginary parts, respectively, of those 
of $D$. This allows to restrict the spectrum of $D$ to a one-dimensional set
by selecting an appropriate function $\F(u,v)$ and requiring $\F(u,v)=0$.
To allow for the eigenvalue $0$ of $D$ the specified curve must go through
zero. In addition, to admit a nonzero index, because of the sum rule \re{res},
it must meet the real axis at least at one further point. Thus, considered
as a function of real arguments, $\F$ must have the properties
\be
\F(0,0)=0 \, , \,\,\, \F(\beta,0)=0 \mbox{ for some } \beta \ne 0 \, \, .
\label{FF}
\ee

Among the many possibilities allowed by \re{FF} there is also 
$\F(u,v) = (u-\rho)^2 + v^2 - \rho^2$ which leads to the (simple form of) 
the GW relation. In this case $\F(u,v)=0$ inserting \re{Duv} reads 
$\rho(D+D\dg)=D\dg D$ which with \re{Ddg}, i.e.~by $\ga$-hermiticity, becomes 
\be
\{\ga,D\}= \rho^{-1} D \ga D   \, \, .
\label{gw}
\ee
In contrast to the original form \cite{gi82}, however, no further 
operator is sandwiched into the r.h.s.~of \re{gw}. This would spoil the 
normality of $D$ and thus also its chiral properties.

\section{WARD IDENTITIES}

Fermionic Ward identities arise from  the condition that 
$\int \df \e^{-S_\f} \cao $ must not change under the transformation  
$\psi'= \exp(i\eta\G)\psi$, $\bar{\psi}'= \bar{\psi} \exp(i\eta\bar{\G})$, 
which leads to 
\ba
\label{w1}
i\int\df \e^{-S_\f}\Big(-\mbox{Tr}(\bar{\G}+\G)\cao\mbox{\hspace*{20mm}}& & \\
-\bar{\psi}(\bar{\G}M+M\G)\psi\cao 
+ \bar{\psi}\bG\frac{\pt \cao}{\pt \bar{\psi}} -
\frac{\pt \cao}{\pt \psi}\G\psi \Big)=0 & &\nonumber 
\ea 
with three contributions, one from the derivative of the integration
measure, one from that of the action, and one from that of $\cao$. 
To proceed properly in the presence of zero modes of $D$ one has to put 
$M=D+\varepsilon$ so that $S_\f=\bar{\psi}M\psi$ and to let $\varepsilon$ 
go to zero in the final result. 

Integrating out the $\bar{\psi}$ and $\psi$ fields in the second term of 
\re{w1} and using general properties of Grassmann variables we obtain 
\ba
iW \int\df \e^{-S_\f}\cao =0\mbox{\hspace*{9mm}with\hspace*{10mm} }
\nonumber& &\\
W = \mbox{Tr}\Big(-\bar{\G}-\G + M^{-1}(\bar{\G}M+M\G)\Big)\,.& & 
\label{w}
\ea
By \re{w} the expectation value in a background gauge field factorizes.
Thus instead of $\,W\langle \cao\rangle_{\f}=0$ it suffices to consider 
$W=0$ in the following.

For the global chiral transformation, which is given by $\G=\bar{\G}=\ga$,
the measure contribution $-\mbox{Tr}(\bG+\G)$ vanishes and one obtains 
\be
W=\mbox{Tr}(M^{-1}\{\ga,M\})
\label{WM}
\ee
so that with $M=D+\varepsilon$ inserted $W/2$ becomes 
\be
\mbox{Tr}\Big((D+\varepsilon)^{-1}\ga D\Big) + 
\varepsilon\mbox{Tr}\Big((D+\varepsilon)^{-1}\ga\Big)\, \,.
\label{gwa2} 
\ee
In the continuum limit the first term in \re{gwa2} gives the topological 
charge as shown some time ago \cite{ke81} for the Wilson-Dirac 
operator and recently \cite{ad98} for the Neuberger operator \cite{ne98}. 
If $D$ is normal and $\ga$-hermitean one has \re{re0} and \re{re1} 
showing that the last term in \re{gwa2} is related to 
the index of $D$ and that the sum of the terms there gives 
\be
W/2 \rightarrow\sum_{\la\mbox{ \scriptsize  real}}\Big(N_+(\la)-N_-(\la)\Big)
\label{was}
\ee
for $\varepsilon \rightarrow 0$. Thus it turns out that one gets just the 
sum rule for real modes \re{res}. 

In case that the GW relation \re{gw} holds the first term in \re{gwa2} can be 
written as  $(2\rho)^{-1}\mbox{Tr}(\ga D)$.  With real eigenvalues only at 
0 and $2\rho$ then only $2\rho\,(N_+(2\rho)-N_-(2\rho))$ 
remains on the r.h.s. of \re{rel} and the sum rule \re{res} simplifies to 
$N_+(0) - N_-(0) + N_+(2\rho) - N_-(2\rho) = 0$. Combining these equations
one gets for the index (in that special case only) the relation 
\be
N_-(0) - N_+(0)=(2\rho)^{-1}\mbox{Tr}(\ga D) \,\,.
\label{ind}
\ee

For the alternative chiral transformation of L\"uscher \cite{lu98} we have
$\G=\ga(1-(2\rho)^{-1}M)$, $\bar{\G}=(1-(2\rho)^{-1}M)\ga$, giving
the measure contribution $+\rho^{-1}\mbox{Tr}(\ga M)$ and the action
contribution $\mbox{Tr}(M^{-1}\{\ga,M\})-\rho^{-1}\mbox{Tr}(\ga M)$.
Thus again the result \re{WM} is obtained, obviously even without 
assuming the GW relation. It is important to realize here that 
in the quantum case with zero modes, because of the necessity of the 
$D+\varepsilon$ regularization, the action is no longer invariant 
with respect to this transformation (if this is not observed as in 
\cite{lu98} the last term in \re{gwa2} gets lost).

The local chiral transformation is given by 
\be
\G=\bar{\G}=\ga\hat{e}(n) \,,\,\,
	 \Big(\hat{e}(n)\Big)_{n''n'}=\delta_{n''n}\delta_{nn'} 
\label{t5}
\ee
for which one obtains
\be
W= \mbox{Tr}\Big((M^{-1}\{\ga\hat{e}(n),M\}\Big) \, \, .
\label{WMe}
\ee
Decomposing $M$ in $\{\ga\hat{e}(n),M\}$ into parts anticommuting and 
commuting with $\ga$ and inserting $M=D+\varepsilon$, \re{WMe} splits
into terms corresponding to the divergence of the singlet axial vector 
current, to the topological-charge density \cite{ke81,ad98}, 
and to the local version of the index.  The local transformation 
related to the above alternative one can be introduced by 
$\G=\ga\hat{e}(n)(1-(2\rho)^{-1}M)\,$, $\bar{\G}
=(1-(2\rho)^{-1}M)\ga\hat{e}(n)$ and leads again to the result \re{WMe}.

\section{GETTING $D$ FROM $H$}

To avoid doublers so far one has to rely on the Wilson-Dirac operator $X/a$,
which is $\ga$-hermitean, however, (with a gauge field) not normal. 
On the other hand, the operator $H=\ga X$ is even hermitean, which
has the consequence that functions of $H$ are well defined, for example 
$E(H)$ with $H\phi_l=\alpha_l\phi_l$ by
$ E(H) = \sum_l E(\alpha_l) \phi_l \phi_l\dg$ where $E(\alpha)$ is a real 
function of real $\alpha$. 

This suggests to use a function of $H$ to get a normal Dirac operator 
$D$. Therefore instead of $X=\ga H$ we consider
\be
D=\ga E(H)+C
\label{YE}
\ee
with a function $E(H)$ and a constant $C$. 
Obviously $E(H)$ must be hermitean and $C$ real in order that $D$ gets 
$\ga$-hermitean. Requiring normality of $D$ we obtain the condition 
\be
[\ga,E(H)^2]=0   \, \, .
\label{5E2}
\ee

Because $H$ does not commute with $\ga$ to satisfy \re{5E2} one 
must require $E(H)^2$ to be independent of $H$. In 
$E(H)^2 = \sum_l E(\alpha_l)^2 \phi_l \phi_l\dg$ this means that 
$E(\alpha_l)^2$ should be constant, i.e.~that 
\be 
E(H)^2=\rho^2\Id \,,\;\; E(\alpha)^2=\rho^2 ,\;\; E(\alpha)=\pm\rho \,.
\label{const}
\ee
By \re{const} the spectrum of $\ga E(H)$ is on a circle with radius $\rho$ 
and center at zero. Thus putting $C=\rho$ the spectrum of $D$ gets the 
correct position. Then $D$ satisfies the GW relation \re{gw}.

Requiring $E(\alpha)$ to be nondecreasing and odd (to keep the behavior of 
$E(H)$ as close as possible to that of $H$) from \re{const} one gets 
$E(\alpha) = \rho \,\epsilon(\alpha)$ with $\epsilon(\alpha)=\pm 1$ for 
$\alpha{>\atop <} 0$. Thus if all $\alpha_l\ne 0$ one arrives just at
the Neuberger operator \cite{ne98}.

If $\alpha_l=0$ occur, $\epsilon(0)$ is also to be specified. Because of 
\re{const} only $+1$ or $-1$ are available for this. To prefer none of these 
choices performing independent calculations of
\be
D=\rho\,(1+\ga\epsilon(H))
\label{DV}
\ee
for each of them and to take the mean of the final results appears 
appropriate.   To show that this is also natural from the point of view of 
the counting of eigenvalue flows of $H$ (introduced in \cite{na93}) we note 
that in the present special case \re{ind} holds and by inserting \re{DV} 
gets $N_-(0) - N_+(0)=\frac{1}{2}\mbox{Tr}(\epsilon(H))$.
In the absence of zero eigenvalues of $H$, in terms of numbers of positive 
and negative eigenvalues of $H$, this becomes 
\be
N_-(0) - N_+(0)= \frac{1}{2} (N_+^H - N_-^H) \,\,.
\label{reNN}
\ee
We now observe that to use \re{reNN} as it is also if zero eigenvalues of $H$ 
occur is adequate. In fact, considering eigenvalue flows as a function of 
the mass parameter, one gets a change of the index by $\frac{1}{2}$ up to the
moment of crossing and a further change by $\frac{1}{2}$ after it. At the very 
moment of crossing the index change in \re{reNN} has reached $\frac{1}{2}$, 
which is in agreement with the respective result of the proposed 
procedure of dealing with $\epsilon(0)$.

In the present special case it is easy to reveal a further feature. Because 
of $\rho\sim 1/a$ the radius of the circle with the spectrum of $D$ increases 
for decreasing lattice spacing $a$. The stereographic projection of this circle 
onto the sphere of complex numbers then approaches the circle through $\infty$ 
on this sphere which is the image of the imaginary axis in the plane. This 
suggests that in the continuum limit, with the spectrum on the imaginary axis, 
the sum rule for chiral differences \re{res} is satisfied by contributions at 
eigenvalues $0$ and $\infty$. Clearly this deserves further investigation and 
should also be observed in cases where different spectral constraints
are used.

\vspace*{2mm}
I am grateful to M.~M\"uller-Preussker and his group for their kind hospitality. 


\begin{thebibliography}{9}
\bibitem{gi82}  P.H. Ginsparg and K.G. Wilson, 
              Phys. Rev. D 25 (1982) 2649.
\bibitem{lu98}  M. L\"uscher,
              Phys. Lett. B 428 (1998) 342. 
\bibitem{ke81}  W. Kerler, 
              Phys. Rev. D 23 (1981) 2384; {\it ibid} 24 (1981) 1595;
                E. Seiler and I.O. Stamatescu,
              Phys. Rev. D 25 (1982) 2177; {\it ibid} 26 (1982) 534 (E).
\bibitem{ad98}  D.H. Adams, hep-lat/9812003; 
                H. Suzuki, hep-th/9812019.
\bibitem{ne98}  H. Neuberger, 
              Phys. Lett. B 417 (1998) 141; 
              {\it ibid} 427 (1998) 353. 
\bibitem{na93}  R. Narayanan and H. Neuberger, 
	      Nucl. Phys. B 443 (1995) 305. 

\end{thebibliography}
\end{document}